\documentclass[aps,prl,twocolumn,superscriptaddress,nofootinbib]{revtex4-2}
\usepackage{amsmath,amssymb,amsthm}
\usepackage{braket}
\usepackage{booktabs}
\usepackage{hyperref}
\usepackage{tikz}
\usetikzlibrary{decorations.pathmorphing,decorations.markings}

\newtheorem{proposition}{Proposition}

\newtheorem{theorem}{Theorem}
\theoremstyle{definition}

\newcommand{\F}{\mathbb F}
\newcommand{\Z}{\mathbb Z}
\newcommand{\Q}{\mathbb Q}
\newcommand{\wt}{\operatorname{wt}}
\newcommand{\one}{\mathbf 1}
\newcommand{\Span}{\operatorname{span}}
\newcommand{\Tr}{\operatorname{Tr}}

\begin{document}

\title{One generation of the Standard Model as a quantum error-correcting code}

\author{Shiroman Prakash}
\email{sprakash@dei.ac.in}
\affiliation{Department of Physics and Computer Science, Dayalbagh Educational Institute, Agra-282005, India}

\author{K. L. Ramya}
\email{ramya144401@dei.ac.in}
\affiliation{Department of Physics and Computer Science, Dayalbagh Educational Institute, Agra-282005, India}

\begin{abstract}
Together with a right-handed neutrino, the fermions of one generation of the Standard Model are labelled by five bits, a combinatorial structure tied to anomaly cancellation. We make this precise. Anomaly cancellation is the vanishing of a cubic form, the defining property of triorthogonal codes, the quantum error-correcting codes behind magic state distillation. Removing the right-handed neutrino gives a $[[15,1,3]]_p$ code for every prime $p$, including the Bravyi--Kitaev code for $p = 2$. For qutrits it is the unique shortest distance-three triorthogonal code.
\end{abstract}

\maketitle

\paragraph*{Introduction.}
The fermions of one generation of the Standard Model have an intriguing combinatorial structure.
Together with a right-handed neutrino, the sixteen fields are labelled by strings of five bits with
an even number of ones. In $SO(10)$ grand unification they are the states of five fermionic
oscillators with an even number
occupied~\cite{Georgi1975,FritzschMinkowski1975,GeorgiGlashow1974,Baez:2009dj}. Colour,
isospin and hypercharge can be read off from the occupation numbers. That the generation is free of
gauge anomalies is well understood group-theoretically, since $SO(10)$ has no cubic
Casimir~\cite{Slansky1981}. We show that this cancellation has a second meaning: it makes the
generation a quantum error-correcting code, of the kind used to distill magic states for
fault-tolerant quantum computation.

Anomaly cancellation and the defining property of these codes are the same cubic equation. Writing
each bit $b$ as a sign $(-1)^b$ turns the sixteen fields into sixteen integer charge vectors. Every
cubic anomaly, a sum of triangle diagrams, is then a value of a cubic form on the lattice they
define~\cite{Adler1969,BellJackiw1969,GrossJackiw1972}, and anomaly cancellation is the vanishing of
this form. Imposed modulo a prime $p$, the same condition defines triorthogonal codes, a class of
quantum error-correcting codes with a transversal non-Clifford gate on which magic state
distillation is built~\cite{MSD,knill2004faulttolerant,Bravyi_2012}. Removing the right-handed
neutrino gives a $[[15,1,3]]_p$ code for every prime $p$. For $p=2$ it is the fifteen-qubit code of
Bravyi and Kitaev~\cite{MSD}; for odd $p$ it is a new code on $p$-level qudits.

Every ingredient of the code has a Standard Model meaning. The physical
qudits are the fifteen quarks and leptons, and the removed right-handed neutrino $\nu^c$ is the
logical qudit. The stabilizers are the symmetries under which $\nu^c$ is neutral. Since $\nu^c$ is a gauge
singlet, they include the gauge charges, colour, weak isospin and hypercharge, and for $p\ge5$
they are exactly these. Their
triorthogonality is Standard Model anomaly cancellation. The logical operator is the one remaining
charge of $SO(10)$. Its mixed anomalies with the gauge charges
vanish, as triorthogonality requires, and its cubic anomaly, which $\nu^c$ would cancel, is the
logical gate. The minimum-weight logical errors are charge-conserving four-fermion operators, such
as the product of Dirac masses $(\nu\nu^c)(ee^c)$, and the construction of the code by
doubling~\cite{Sullivan2024,PrakashCatalogue} is the pairing of quarks and leptons with their Dirac
partners.

For qutrits the code is the shortest of its kind and unique: an exhaustive search finds no shorter triorthogonal qutrit code of distance at least three, and none of length fifteen other than this one, up to equivalence. The shortest previously known had nineteen qutrits~\cite{PrakashCyclic}. We comment on its potential for distillation below.

Background for readers from either community, proofs and further results are in the Supplemental Material.

\paragraph*{From anomalies to codes.}
Consider $n$ Weyl fermions $\psi_\ell$, $\ell=0,\dots,n-1$, with chiralities $c_\ell=\pm1$ and integer
charge vectors $q_\ell\in\Z^r$, with components $q_{\ell,i}$, under a torus $U(1)^r$. A $U(1)$
subgroup, labelled by $t\in\Z^r$, gives the fermion $\psi_\ell$ the charge $x_\ell=q_\ell\cdot t$.
Each symmetry thus has a vector
$x=(x_0,\dots,x_{n-1})\in\Z^n$ of charges, and these vectors form a lattice $\Lambda\subseteq\Z^n$ with
an $r\times n$ generator matrix, whose columns are the $q_\ell$ and whose rows list the charges under
the $r$ factors. The mixed cubic anomaly of three symmetries, with
vectors $x$, $y$ and $z$, is the sum of triangle diagrams
\begin{equation}\label{eq:tau}
\tau_c(x,y,z)=\sum_\ell c_\ell x_\ell y_\ell z_\ell.
\end{equation}
The spectrum is free of cubic anomalies exactly when $\tau_c$
vanishes on $\Lambda$, and we call such a lattice \emph{triorthogonal}. 

Imposed modulo a prime $p$, the same condition defines triorthogonal
codes~\cite{Bravyi_2012,Nezami_2022,PrakashCyclic,Saha:2025frb}. A subspace $W\subseteq\F_p^n$ on which $\tau_c$
vanishes, for coefficients $c\in(\F_p^\times)^n$, is a \emph{triorthogonal space}.\footnote{Unlike an earlier definition~\cite{Krishna_2019}, we do not require self-orthogonality,
which the transversal gate does not need~\cite{PrakashCyclic,SahaNotes}; our code is not
self-orthogonal for any odd $p$.} Deleting
coordinate $0$ turns it into a CSS code on the other $n-1$ coordinates. Its $X$-stabilizers are the
vectors of $W$ that vanish at coordinate $0$, and its logical $X$ is a vector $g\in W$ with $g_0=1$.
Every triorthogonal code encoding one qudit arises in this way.

Triorthogonality enables a transversal non-Clifford gate on the encoded qudit. The gate is
$\bigotimes_{\ell\neq0}U^{c_\ell}$, where $U$ is the $\pi/8$ gate. For $p=2$,
$U=T=\operatorname{diag}(1,e^{i\pi/4})$, and for $p>2$, its analogue~\cite{HowardVala,Cui} is
\[
U=\begin{cases}
\sum_j\omega^{j^3}\ket j\bra j,\quad \omega=e^{2\pi i/p}, & p\ge5,\\
\operatorname{diag}(1,\zeta,\zeta^{-1}),\quad \zeta=e^{2\pi i/9}, & p=3.
\end{cases}
\]
The logical states are $\ket{\bar a}\propto\sum_x\ket{ag+x}$, summed over the stabilizers $x$. For $p\ge5$, the basis state labelled by
$w=ag+x\in W$, for which $w_0=a$, picks up one phase $\omega^{c_\ell w_\ell^3}$ from each qudit, and since
$\tau_c(w,w,w)=\sum_\ell c_\ell w_\ell^3=0$,
\begin{equation}\label{eq:gate}
\Big(\bigotimes_{\ell\neq0}U^{c_\ell}\Big)\ket{w}=\omega^{\sum_{\ell\neq0}c_\ell w_\ell^3}\ket{w}
=\omega^{-c_0a^3}\ket{w}.
\end{equation}
The phase is the same for every term of $\ket{\bar a}$, so the gate acts as logical $U^{-c_0}$. For $p=3$ and $p=2$ the phases
of $U$ are ninth and eighth roots of unity, while the code is defined only modulo $3$ and $2$, so the same argument fixes the gate up to a Pauli
correction for $p=3$~\cite{PrakashCyclic} and a Clifford correction for $p=2$~\cite{Bravyi_2012}.

Reducing a triorthogonal lattice modulo $p$ gives a triorthogonal space,
$W=\Lambda\bmod p$. In physical terms, a vector $w\in W$ lists the charges of the fermions, modulo $p$, under a $\Z_p$ subgroup of
the torus. Deleting coordinate $0$ means
removing a fermion, $\psi_0$, and the remaining fermions are the physical qudits. The $X$-stabilizers
are then the $\Z_p$ symmetries under which $\psi_0$ is neutral, and the logical $X$ is one under which
$\psi_0$ has charge $1$. The exponent $\sum_{\ell\neq0}c_\ell w_\ell^3$ in Eq.~\eqref{eq:gate} is the
cubic anomaly of the symmetry $w$ on the remaining fermions, and it is fixed because the full spectrum is anomaly
free. The logical gate is thus the anomaly that the removed fermion would cancel.

The distance relevant for distillation is the $Z$-distance $d$, the smallest Hamming weight of an
undetectable $Z$-error that acts nontrivially on the encoded qudit. Such an error $Z(v)$ has $v$
orthogonal to every $X$-stabilizer but not to $g$, so $\sum_{\ell\neq0}v_\ell h_\ell$ is a nonzero
multiple of $h_0$, where the $h_\ell$ are the columns of a generator matrix of $W$. The error is thus
a linear relation expressing the charge vector of $\psi_0$ through those of the fermions in its
support $J$, and
\begin{equation}\label{eq:distance}
d=\min\{|J|:\ 0\notin J,\ h_0\in\Span\{h_\ell :\ell\in J\}\}.
\end{equation}Physically, a logical error is a combination of fields containing $\psi_0$ whose total charge vanishes
modulo $p$, and $d+1$ is the smallest number of fields in such a combination. A vectorlike partner of $\psi_0$ has charge vector $-q_0$, and an identical copy has $q_0$; in
either case $h_0$ lies in the span of a single column and $d=1$. Useful triorthogonal codes
therefore need spectra that are both anomaly free and chiral.  Because $\Lambda$ is defined over $\Z$, one spectrum gives a code for every
odd prime $p$ with $q_0\not\equiv0\pmod p$, with the same distance as over $\Q$ for all but finitely
many $p$.

We summarize the construction for odd primes in the following proposition.

\begin{proposition}\label{prop:dict}
Let $\psi_0,\dots,\psi_{n-1}$ be Weyl fermions with chiralities $c_\ell$ and charge vectors
$q_\ell\in\Z^r$, free of cubic anomalies, and let $\Lambda$ be their charge lattice. Let $p$ be an odd
prime with $q_0\not\equiv0\pmod p$. Then $W=\Lambda\bmod p$ is a triorthogonal space for the
coefficients $c\bmod p$, and removing $\psi_0$ gives a triorthogonal code on the other $n-1$
fermions. The transversal gate
$\bigotimes_{\ell\neq0}U^{c_\ell}$ implements logical $U^{-c_0}$ exactly.
\end{proposition}

\begin{proof}
The identity $\tau_c=0$ on $\Lambda$ holds over $\Z$, and so modulo every $p$. For $p\ge5$ the gate
follows from Eq.~\eqref{eq:gate}. For $p=3$ the phases of $U$ are powers of $\zeta$, and the exponent
must be computed modulo $9$. Since $(x+3k)^3\equiv x^3\pmod9$, it depends only on residues modulo $3$.
A codeword $x$ with $x_0=a$ is the reduction of some $y\in\Lambda$, so
$\sum_{\ell\neq0}c_\ell x_\ell^3\equiv\sum_{\ell\neq0}c_\ell y_\ell^3=-c_0y_0^3\equiv-c_0a^3\pmod9$, the same
for every term of $\ket{\bar a}$.
\end{proof}

The above construction becomes degenerate for $p=2$, so qubits need special treatment. In addition, $T$ depends on Hamming weights modulo $8$, which $\Lambda\bmod2$ does not control. When
all charges are $\pm1$ we use instead their sign bits $(1-q_{\ell,i})/2$, which together with $\one$
span a triply even code if, besides the gauge anomalies, the anomalies $\Tr F^3$ and $\Tr(FQ_iQ_j)$,
$i\neq j$, of fermion number $F$ vanish modulo $16$. For one generation they do, as we check below.

\paragraph*{One generation and its anomalies.}
We now specialize to one generation of the Standard Model, embedded in $SO(10)$. For a nonabelian
gauge group it is enough to work with its Cartan torus, a maximal set of commuting $U(1)$ factors.
Every generator is conjugate into the Cartan algebra, so the cubic anomalies vanish exactly when they
vanish on the torus. For $SO(10)$ the torus is $U(1)^5$, and each component of a multiplet is a fermion
whose charge vector is its weight. The spinor $\mathbf{16}$ is the Fock space of five fermionic
oscillators with an even number occupied, and its weights, scaled by $2$, are the sign vectors
$q\in\{\pm1\}^5$ with an even number of minus signs, where $q_i=-1$ when oscillator $i$ is occupied.
The states with zero, two and four oscillators occupied form the $\mathbf 1$, $\mathbf{10}$ and
$\bar{\mathbf 5}$ of $SU(5)$, on which $\chi=\sum_iq_i$ takes the values $5$, $1$ and $-3$. Taking
oscillators $1$--$3$ as colour and $4$--$5$ as weak, a state with $n_c$ colour and $n_w$ weak
oscillators occupied has hypercharge $Y=n_w/2-n_c/3$. The empty state, with weight
$\one=(1,1,1,1,1)$, is the conjugate right-handed neutrino $\nu^c$, and the other fifteen states are
the quarks and leptons of one generation, all written as left-handed fields. The generation is chiral -- flipping the sign of $q$ changes the number of occupied oscillators from $j$ to $5-j$, of the opposite
parity, so no weight of the $\mathbf{16}$ has its negative in the $\mathbf{16}$.

In the notation of the previous section, the spectrum consists of $n=16$ left-handed fermions, so
$c_\ell=1$, with $r=5$ and the weights $q$ as charge vectors. The cubic anomaly of the $U(1)$ labelled
by $t\in\Z^5$ is $\mathcal A(t)=\sum_q(q\cdot t)^3$, summed over the sixteen weights. Expanding the cube gives the sums $\sum_qq_iq_jq_k$, and since $q_i^2=1$,
each reduces to $\sum_qq_i$ or to $\sum_qq_iq_jq_k$ with distinct indices. Both vanish, because on the sixteen weights each of
$q_i$ and $q_iq_jq_k$ equals $+1$ and $-1$ equally often, so the charge lattice $\Lambda$ is triorthogonal -- every cubic anomaly of the
generation, including mixed anomalies, vanishes. The quadratic sums are $\sum_qq_iq_j=16\,\delta_{ij}$, so the fermion-number anomalies that the
qubit construction needs, $\Tr F^3=n=16$ and $\Tr(FQ_iQ_j)=0$ for $i\neq j$, vanish modulo $16$.

We remove $\psi_0=\nu^c$, whose weight is $q_0=\one$. A generator matrix of the resulting code, with
the logical row $q_1$ above the Cartan generators $(q_i-q_{i+1})/2$ of $\mathfrak{su}(5)$ as
stabilizers, is
\begin{equation}\label{eq:H}
\setlength{\arraycolsep}{2.5pt}
\left(\begin{array}{cccccccccc|ccccc}
\bar1 & \bar1 & \bar1 & \bar1 & 1 & 1 & 1 & 1 & 1 & 1 & 1 & \bar1 & \bar1 & \bar1 & \bar1\\ \hline
0 & \bar1 & \bar1 & \bar1 & 1 & 1 & 1 & 0 & 0 & 0 & 1 & \bar1 & 0 & 0 & 0\\
\bar1 & 1 & 0 & 0 & 0 & \bar1 & \bar1 & 1 & 1 & 0 & 0 & 1 & \bar1 & 0 & 0\\
0 & \bar1 & 1 & 0 & \bar1 & 1 & 0 & 0 & \bar1 & 1 & 0 & 0 & 1 & \bar1 & 0\\
0 & 0 & \bar1 & 1 & 0 & \bar1 & 1 & \bar1 & 1 & 0 & 0 & 0 & 0 & 1 & \bar1
\end{array}\right),
\end{equation}
with $\bar1=-1$. Its first ten columns are the fields of the $\mathbf{10}$,  and its last five are those of the $\bar{\mathbf 5}$.

\begin{theorem}\label{thm:sm}
For every prime $p$, removing $\nu^c$ from one generation gives a triorthogonal $[[15,1,3]]_p$ code on
the fifteen Standard Model fermions, on which $U^{\otimes15}$ implements logical
$U^\dagger$ exactly. For $p=2$ it is the fifteen-qubit code of Bravyi and Kitaev~\cite{MSD}.
\end{theorem}

\begin{figure}[t]
\centering
\begin{tikzpicture}[x=0.86cm,y=1.0cm,
  f/.style={circle,draw=black,fill=white,inner sep=1.7pt},
  r/.style={circle,draw=red!75!black,fill=red!75!black,inner sep=1.7pt},
  b/.style={circle,draw=blue!70!black,fill=blue!70!black,inner sep=1.7pt},
  lab/.style={font=\scriptsize,text height=1.6ex,text depth=0.5ex}]
\foreach \i/\t/\bt/\st/\sb in {1/u^c_1/u_1/r/f,2/u^c_2/u_2/b/b,3/u^c_3/u_3/f/f,4/d^c_1/d_1/r/f,
                                5/d^c_2/d_2/f/f,6/d^c_3/d_3/f/f,7/e^c/e/r/f}{
  \draw[gray] (\i,0)--(\i,1);
  \node[\st] (t\i) at (\i,1) {}; \node[lab,above=0pt] at (t\i.north) {$\t$};
  \node[\sb] (b\i) at (\i,0) {}; \node[lab,below=0pt] at (b\i.south) {$\bt$};}
\draw[gray,dashed] (8.4,0)--(8.4,1);
\node[circle,draw=black,dashed,thick,minimum size=7pt,inner sep=0pt] (nc) at (8.4,1) {};
\node[lab,above=0pt] at (nc.north) {$\nu^c$};
\node[b] (nu) at (8.4,0) {}; \node[lab,below=0pt] at (nu.south) {$\nu$};
\draw[gray!70,densely dotted] (7.7,-0.55)--(7.7,1.55);
\node[lab,anchor=east] at (0.55,1) {$f^c$};
\node[lab,anchor=east] at (0.55,0) {$f$};
\node[lab,anchor=west] at (8.65,1) {logical};
\node[lab,anchor=west] at (8.65,0) {seed};
\end{tikzpicture}
\caption{The fifteen-qudit code as a doubled code. Vertical lines join the right-handed fields $f^c$
(top), one copy of a $[[7,1,3]]^P_p$ code with $\mu=2T_{3R}$, to their Dirac partners $f$ (bottom). The
seed is $\nu$; its partner $\nu^c$ (dashed) is the logical qudit. Red and blue mark the
minimum-weight logicals $\nu^c=e^c+u^c_1-d^c_1$, within one copy, and $\nu^c=u_2+u^c_2-\nu$, through
the seed.}
\label{fig:code}
\end{figure}
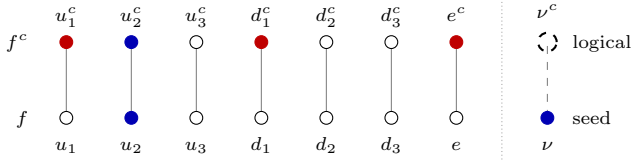

\begin{proof}
For odd $p$, triorthogonality and the exact gate follow from Proposition~\ref{prop:dict}, because all sixteen fields are left-handed and $q_0=\one$ is nonzero modulo every $p$. For the distance, the relation $\nu^c=e^c+u^c_1-d^c_1$ among
weights (Fig.~\ref{fig:code}) gives $d\le3$. Conversely, no one or two of the other weights span that of $\nu^c$ over $\Q$
(Proposition~\ref{prop:spinor}). Such a span could appear modulo $p$ only if the rank of the
weights involved dropped, that is, only if $p$ divided a nonzero $2\times2$ or $3\times3$ minor
of their $\pm1$ entries. These minors are $\pm2$ or
$\pm4$, which no odd prime divides, so $d=3$ for every odd $p$.

For $p=2$, the sign bits of the sixteen weights, together with $\one$, span the first-order
Reed--Muller code $RM(1,4)$, which is triply even. Deleting the coordinate of $\nu^c$ gives the
fifteen-qubit code, whose distance is $3$, and on which $T^{\otimes15}$ implements logical $T^\dagger$.
\end{proof}

\paragraph*{Physical interpretation of the code.}
Table~\ref{tab:dict} summarizes the dictionary. The stabilizers are the Cartan generators of
$SU(5)$, which over $\Q$ span those of the Standard Model gauge group, so their triorthogonality is
anomaly cancellation in the form $A(\mathbf{10})+A(\bar{\mathbf 5})=0$~\cite{GeorgiGlashow1974}.
The logical operator can be taken to be $B-L$, or equivalently $\chi/5$, since the two differ by
stabilizers. The anomaly of $U(1)_\chi$ on the fifteen fermions is
$\sum\chi^3=10\cdot1^3+5\cdot(-3)^3=-125=-\chi(\nu^c)^3$, as the general argument requires.

The minimum-weight logical errors have $25$ supports. Each is a
Cartan-neutral combination of $\nu^c$ with three other fields, all with coefficients $\pm1$. They are four-fermion monomials, such as
the product of Dirac masses $(\nu\nu^c)(ee^c)$, the right-handed current of Fig.~\ref{fig:code}, and the
$B-L$-conserving $\nu^cd^cd^cu^c$. They are not Standard Model invariants by themselves, but each has the charge structure of an
operator generated at tree level by exchange of a $\mathbf{10}$ of $SO(10)$.

The code is the $\mathbf{16}$ of $SO(10)$ with one weight
removed, and the Standard Model enters through our choice of which weight to remove. Because the Weyl group is transitive, removing any of the sixteen fields gives an
equivalent code. Gauge symmetry singles out $\nu^c$. Requiring the Cartan
generators of $SU(3)_c\times SU(2)_L\times U(1)_Y$ to be stabilizers, over $\Z$ or modulo any
$p\ge5$, forces the removed field to be $\nu^c$. Modulo $3$ the positron $e^c$ also satisfies this requirement, because
its hypercharge in integral units is $6Y=6\equiv0\pmod3$. It is the singlet of flipped
$SU(5)$~\cite{Barr1982,Derendinger1984}.

\begin{table}[t]
\centering\small
\begin{tabular}{@{}p{0.45\columnwidth}p{0.49\columnwidth}@{}}
\toprule
\raggedright triorthogonal code & \raggedright chiral gauge theory\tabularnewline
\midrule
\raggedright physical qudit & \raggedright Weyl fermion $\psi_\ell$\tabularnewline
\raggedright logical qudit & \raggedright removed fermion $\psi_0$\tabularnewline
\raggedright coefficient $c_\ell$ & \raggedright chirality $\pm1$\tabularnewline
\raggedright triorthogonal space & \raggedright anomaly-free spectrum\tabularnewline
\raggedright $X$-stabilizer & \raggedright symmetry under which $\psi_0$ is neutral\tabularnewline
\raggedright logical $X$ & \raggedright symmetry under which $\psi_0$ has charge $1$\tabularnewline
\raggedright logical gate $U^{-c_0}$ & \raggedright cubic anomaly that $\psi_0$ would cancel\tabularnewline
\raggedright minimum-weight logical & \raggedright neutral set of $d+1$ fermions containing $\psi_0$\tabularnewline
\raggedright automorphisms & \raggedright Weyl group fixing $\psi_0$\tabularnewline
\raggedright doubled copies & \raggedright Dirac pairs $(f,f^c)$\tabularnewline
\raggedright linking row & \raggedright weak isospin $2T_{3L}$\tabularnewline
\bottomrule
\end{tabular}
\caption{The dictionary. The last three rows hold for the code of Theorems~\ref{thm:sm}
and~\ref{thm:dirac}.}
\label{tab:dict}
\end{table}

\paragraph*{Doubling and Dirac pairing.}
Doubling~\cite{Betsumiya_2012,BravyiCross2015,Sullivan2024} is the standard way to build small
triorthogonal codes of growing distance~\cite{Jain_2025}. It extends to odd $p$~\cite{PrakashCatalogue}. Its inputs are a
triorthogonal seed $\mathcal T$, with logical exponent $\lambda_{\mathcal T}$ and distance
$d_{\mathcal T}$, and a $P$-code $(\mathcal S,\mu)$, a space $\mathcal S$ with coefficients $\mu$
such that $\sum_\ell\mu_\ell s_\ell s'_\ell=\sum_\ell\mu_\ell s_\ell=0$ for all $s,s'\in\mathcal S$ and
$N=\sum_\ell\mu_\ell\neq0$. Such a code carries a transversal phase gate, and for a code of length $n_D$ we
write its parameters as $[[n_D,1,d_{\mathcal S}]]^P_p$. Doubling places two copies of $\mathcal S$ side by side, with every stabilizer of
$\mathcal S$ acting oppositely on paired columns, and connects them to the seed by a single
\emph{linking row} that is nonzero on only one copy. For a scalar $\kappa$ with
$N\kappa^3+\lambda_{\mathcal T}=0$, the result is triorthogonal with distance
$\min(d_{\mathcal S},d_{\mathcal T}+2)$.

Doubling thus needs pairs of qudits on which the stabilizers are vectorlike, together with one
chiral linking row. A generation has this structure. It is vectorlike under $SU(3)_c\times U(1)_{\rm em}$, because the Weyl group element
$\sigma:q\mapsto(-q_1,-q_2,-q_3,q_4,-q_5)$ of $SO(10)$ maps each $f^c$ to its Dirac partner $f$, with
opposite colour and electric charge, and all of its chirality resides in weak isospin. In terms of the
weights,
\[
T_{3L}=\tfrac{q_4-q_5}4,\;\; T_{3R}=\tfrac{q_4+q_5}4,\;\; B-L=\tfrac{q_1+q_2+q_3}3,
\]
with $Y=\frac12(B-L)-T_{3R}$ and $Q_{\rm em}=T_{3L}+Y$~\cite{PatiSalam1974}.

\begin{theorem}\label{thm:dirac}
After rescaling columns by $\pm1$, the code of Theorem~\ref{thm:sm} is the doubling of the trivial
seed with a $[[7,1,3]]^P_p$ code with coefficients $\pm1$, which for $p=2$ is the Steane code. The
first copy holds $u^c$, $d^c$ and $e^c$, with coefficients $\mu=2T_{3R}$. The second copy holds their
Dirac partners $u$, $d$ and $e$, each in the column paired with its partner. The seed is the neutrino
$\nu$, whose partner was removed. The paired stabilizers are the charges $(q_5-q_i)/2$, $i=1,2,3$,
which over $\Q$ span the Cartan algebra of $SU(3)_c\times U(1)_{\rm em}$, and the linking row is
$2T_{3L}$.
\end{theorem}

\noindent The proof is in the Supplemental Material. 

\paragraph*{Minimality and uniqueness for qutrits.}
The shortest distance-three triorthogonal qutrit code known before had nineteen
qutrits~\cite{PrakashCyclic}. The code from a generation is shorter, and a computer search reveals the following result.

\begin{theorem}[computer-assisted]\label{thm:unique}
No triorthogonal qutrit code with $d\ge3$ has fewer than fifteen qutrits, and every triorthogonal
$[[15,1,3]]_3$ code is equivalent to the code of Theorem~\ref{thm:sm}. Its monomial automorphism
group is $S_5\times\Z_2$, where $\Z_2$ negates every qutrit.
\end{theorem}

\noindent The proof reduces triorthogonality to polynomial conditions on a normal form and searches
them exhaustively. The $S_5$ permutes the five oscillators. It is the
stabilizer of $\nu^c$ in the Weyl group, and its orbits are the $\mathbf{10}$ and the
$\bar{\mathbf 5}$. The smallest distance-three qutrit triorthogonal code thus admits a
description as a generation of the Standard Model, with exactly the symmetry that the generation
keeps, up to an overall sign.\footnote{The fifteen-qubit code, also the shortest distance-three qubit triorthogonal
code~\cite{Nezami_2022,WillsJainSingh2026}, has instead the larger group $GL(4,2)\cong A_8$, which
contains this $S_5$ but is transitive on all fifteen qubits~\cite{macwilliams77}, mixing the
$\mathbf{10}$ with the $\bar{\mathbf 5}$.}

For magic state distillation~\cite{ACB,CampbellAnwarBrowne,campbell2014enhanced,Prakash:2025azi},
the leading output error of the qutrit code at equal input infidelity is
$6.25\,\epsilon^3$, against $35\,\epsilon^3$ for the fifteen-qubit code, since on each of its $25$
weight-three logical supports only $2$ of the $8$ error patterns are logical.

\paragraph*{Outlook.}
For $p=2$, the generation is Bravyi-Kitaev's famous 15-qubit code, a standard benchmark for magic state distillation. The qutrit code is also particularly interesting for its potential applications to distillation. Combined with high-rate qutrit codes~\cite{Prakash:2025azi}, the fifteen-qutrit code could lower the cost of qutrit magic state distillation in some noise
regimes, which we study, in part, elsewhere~\cite{PrakashCatalogue}. 

Doubling preserves integrality, so it can be iterated within
anomaly-free spectra, but only with $P$-codes whose coefficients take both signs. Doubling the
fifteen-qutrit code with the ternary Golay code~\cite{Golay1949,2020golay} gives a $[[37,1,5]]_3$
code~\cite{PrakashCatalogue}, but since the Golay coefficients are all $1$, this doubling does not lift
to an anomaly-free spectrum; an integral $P$-lattice of distance five with $|N|=1$ would give one that
does.

The construction applies beyond $SO(10)$. The larger spinors, of $SO(4k+2)$ with $k\ge3$, give $[[2^{2k}-1,1,3]]_p$
codes for odd $p$ in the same way, and the $\mathbf{27}$ of $E_6$ gives $d=2$. Heterotic compactifications on Calabi--Yau threefolds, which give
chiral $E_6$ and $SO(10)$ spectra~\cite{CHSW1985}, are natural further inputs, provided their
generations are not identical copies.

\begin{acknowledgments}
This work was supported by a DST-SERB grant (CRG/2021/009137) and by the DST-WISE KIRAN Ph.D.
fellowship (grant DST/WISE-PhD/PM/2023/32(G)). We thank the International Centre for
Theoretical Sciences (ICTS) for hospitality during several visits, including the discussion meeting
``A Hundred Years of Quantum Mechanics'' (ICTS/qm100/2025/01) and the programme ``Monopole Moduli''
(ICTS/MM/2026/09), and R.~Loganayagam for discussions. A large language model (Claude Opus 5.5) was used to assist with coding and
drafting; we verified all results and take full responsibility for the content. Scripts
reproducing all numerical results are available in the accompanying repository~\cite{LetterRepo}. SP
thanks Prof.~P.~S.~Satsangi for discussions on qudit magic state distillation some years ago that
inspired this work~\cite{DPS}.
\end{acknowledgments}

\bibliography{qudit2026}

\clearpage
\onecolumngrid
\appendix
\setcounter{proposition}{0}
\renewcommand{\theproposition}{S\arabic{proposition}}
\renewcommand{\theHproposition}{S\arabic{proposition}}

\begin{center}\textbf{\large Supplemental Material}\end{center}

This Supplement has three parts. Part A gives background for readers from either community, lists
the sixteen weights as Standard Model fields (Table~\ref{tab:cols}), and relates this work to previous
literature. Part B proves the
main-text results: the reduction of triorthogonal lattices, the generator matrix and the qubit case, doubling and
Theorem~\ref{thm:dirac}, minimality and uniqueness (Theorem~\ref{thm:unique}), and distillation
performance. Part C interprets the minimal logical errors as four-fermion operators, explains which fermions
can be removed, from $\nu^c$ alone to several at once, and treats other representations, which
completes the proof of Theorem~\ref{thm:sm}. The final section describes the code that
reproduces every statement.

\section*{Part A: Background for readers from either side}

\paragraph{Triorthogonal codes, for physicists.} A qudit has basis states $\ket j$, $j\in\F_p$, on
which the Pauli operators act as $X\ket j=\ket{j+1}$ and $Z\ket j=\omega^j\ket j$, with
$\omega=e^{2\pi i/p}$. Strings of Paulis on $n$ qudits are labelled by vectors of $\F_p^n$, as
$X(x)=\bigotimes_\ell X^{x_\ell}$ and $Z(v)=\bigotimes_\ell Z^{v_\ell}$. A CSS code encoding one qudit is specified
by a subspace $A\subseteq\F_p^n$, whose vectors label the $X$-type stabilizers, and a vector $g\notin A$,
which labels the logical $X$. Its logical basis states are uniform superpositions over the cosets of
$A$,
\[
\ket{\bar a}\propto\sum_{x\in A}\ket{ag+x},\qquad a\in\F_p,
\]
and its $Z$-type stabilizers are the $Z(v)$ with $v$ orthogonal to $A$ and to $g$, which leave every
such state unchanged.

Unitaries that map Pauli operators to Pauli operators under conjugation are \emph{Clifford gates}. For
odd $p$ they include the diagonal gates $\sum_j\omega^{f(j)}\ket j\bra j$ with $f$ a polynomial of
degree at most two. Clifford gates are comparatively easy to implement fault-tolerantly, but circuits
of Clifford gates acting on stabilizer states, with Pauli measurements, can be simulated efficiently
on a classical computer~\cite{Gottesman:1998hu}, so universal quantum computation needs one further, non-Clifford gate. The
Clifford hierarchy~\cite{Gottesman:1999tea} organizes such gates. Its first level is the Pauli group, and level $k+1$ consists
of the unitaries that map Pauli operators, under conjugation, into level $k$. The Clifford gates form
the second level, and the diagonal gate $U$ with cubic phases used below lies in the third. Magic
state distillation, described at the end of this section, supplies such a gate.

Now apply the transversal gate $\bigotimes_\ell U^{c_\ell}$, with $U=\sum_j\omega^{j^3}\ket j\bra j$ for
$p\ge5$ and coefficients $c_\ell\neq0$. It is diagonal, so it multiplies each basis state $\ket{ag+x}$ by a phase,
$\omega$ raised to the power
\[
\sum_\ell c_\ell (ag+x)_\ell^3=a^3\tau_c(g,g,g)+3a^2\tau_c(g,g,x)+3a\,\tau_c(g,x,x)+\tau_c(x,x,x),
\]
where $\tau_c(x,y,z)=\sum_\ell c_\ell x_\ell y_\ell z_\ell$. If the last three terms vanish for every $x\in A$, the
phase is $\omega^{\lambda a^3}$ with $\lambda=\tau_c(g,g,g)$, the same for every term of $\ket{\bar
a}$, so the gate acts on the encoded qudit as $U^\lambda$, a non-Clifford gate when $\lambda\neq0$. Triorthogonality
guarantees this. It demands that $\tau_c$ vanish on every triple of vectors of $A+\F_pg$ containing at
least one vector of $A$. Appending to each vector $ag+x$ a new coordinate equal to $a$, with coefficient
$-\lambda$, turns $A+\F_pg$ into the triorthogonal space $W$ of the main text, and deleting the
coordinate again is puncturing.

For $p=3$ the cube is trivial modulo $3$, since $j^3\equiv j$, so $\omega^{j^3}$ is a Clifford gate.
The non-Clifford qutrit gate uses ninth roots of unity instead,
$U=\operatorname{diag}(1,\zeta,\zeta^{-1})=\sum_j\zeta^{j^3}\ket j\bra j$ with $\zeta=e^{2\pi i/9}$ and
$j\in\{0,1,-1\}$, and the exponent above must then be computed modulo $9$. Triorthogonality modulo $3$
controls the cross terms, whose coefficients are $3$, but fixes the remaining phases only up to
multiples of $3$ in the exponent, which is a Pauli $Z$ correction. When no correction is needed the
gate is \emph{exact}. For codes that lift to an anomaly-free spectrum it always is
(Proposition~\ref{prop:dict}).

In magic state distillation the code is used in reverse~\cite{MSD,Bravyi_2012}. Noisy copies of the
magic state $U\ket+$ are twirled so that each suffers a random error $Z^k$, then decoded, and checked by
measuring the $X$-stabilizers. An error $Z(e)$ is detected unless $e$ is orthogonal to $A$. If it is also orthogonal to $g$, it acts trivially, and otherwise it is a logical error. The protocol therefore
fails only on the $Z$-logicals, the vectors of $A^\perp$ not orthogonal to $g$, and the output error is
dominated by the number of those of minimum Hamming weight. Only the $Z$-distance matters, and for
triorthogonal codes it is never larger than the $X$-distance~\cite{Bravyi_2012,PrakashCyclic}. The
quadratic analogue of triorthogonality, $\sum_\ell c_\ell x_\ell y_\ell=0$, gives a transversal phase gate
$P=\sum_j\omega^{j^2}\ket j\bra j$ in the same way. Codes with this property are the $P$-codes used in doubling.

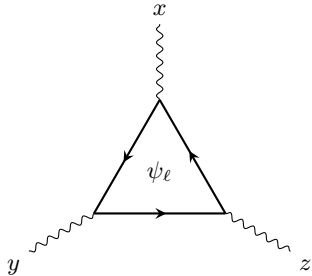
\begin{figure}[h]
\centering
\begin{tikzpicture}[
  fermion/.style={thick,postaction={decorate},
    decoration={markings,mark=at position 0.55 with {\arrow{stealth}}}},
  boson/.style={decorate,decoration={snake,amplitude=1.2pt,segment length=5pt}}]
\coordinate (A) at (0,1);
\coordinate (B) at (-0.87,-0.5);
\coordinate (C) at (0.87,-0.5);
\draw[fermion] (A) -- (B);
\draw[fermion] (B) -- (C);
\draw[fermion] (C) -- (A);
\draw[boson] (A) -- (0,2);
\draw[boson] (B) -- (-1.73,-1);
\draw[boson] (C) -- (1.73,-1);
\node[above] at (0,2) {$x$};
\node[below left] at (-1.73,-1) {$y$};
\node[below right] at (1.73,-1) {$z$};
\node at (0,0.05) {$\psi_\ell$};
\end{tikzpicture}
\caption{The triangle diagram. A Weyl fermion $\psi_\ell$ in the loop couples to three $U(1)$ gauge
fields, under which it has charges $x_\ell$, $y_\ell$ and $z_\ell$, and contributes
$c_\ell x_\ell y_\ell z_\ell$ to their mixed anomaly. Summing over the fermions gives
$\tau_c(x,y,z)$.}
\label{fig:triangle}
\end{figure}

\paragraph{Anomalies, for coding theorists.} The physical input is a list of
fermions and their charges. A Weyl fermion is a two-component spinor field of definite chirality,
left- or right-handed. A right-handed fermion of charge $q$ is equivalent to its conjugate, a
left-handed fermion of charge $-q$, so any spectrum can be written with left-handed fields only. In code terms this negates a column together with its coefficient, which leaves $\tau_c$ unchanged. A
symmetry acting on the fermions by phases is a $U(1)$, and assigns each fermion an integer charge. For
a gauge theory to be consistent, its cubic gauge anomalies must vanish. That is, for every choice of three of its $U(1)$ symmetries, the one-loop triangle diagram of Fig.~\ref{fig:triangle}, to which each fermion contributes the product of
its three charges with the sign of its chirality, must sum to zero. If $x$, $y$ and $z$ are the charge
vectors of the three symmetries, indexed by fermions, and $c$ the vector of chiralities, the anomaly is $\tau_c(x,y,z)$.

The smallest instance of the dictionary has a single $U(1)$. Five
left-handed fermions with charges $(1,5,-7,-8,9)$ have vanishing cubic and linear charge
sums~\cite{CDF2019}, so their charge lattice, of rank one, is triorthogonal. Removing the fermion of
charge $1$ leaves a code on four qudits with no stabilizers, since no nonzero symmetry leaves that
fermion neutral, and with logical $X$ given by the remaining charges $(5,-7,-8,9)$. Its logical exponent
is $\sum x^3=125-343-512+729=-1$, which is minus the cube of the removed charge, that is, the anomaly the removed fermion would cancel. The distance is $1$, since with one $U(1)$ every charged column is proportional to the
removed one. Codes need lattices of higher rank, which nonabelian groups provide.

\paragraph{Nonabelian groups.} For a nonabelian group such as $SU(5)$ or $SO(10)$ one keeps a maximal set of commuting generators, the
Cartan torus $U(1)^r$. Each component of a multiplet is then a fermion with a definite charge vector in
$\Z^r$, a \emph{weight} of the representation. There is no loss of generality when restricting to the torus, because every element of the Lie algebra is conjugate into the Cartan algebra and the anomaly is invariant under conjugation. The Weyl group, a
finite group generated by reflections, permutes the weights of every representation and acts linearly
on the charge lattice, so its elements that fix the removed weight are automorphisms of the code. For
$SO(10)$ the weights of the spinor $\mathbf{16}$, scaled by $2$, are the sign vectors
$q\in\{\pm1\}^5$ with an even number of minus signs, and the Weyl group acts by permuting the five
coordinates and by changing an even number of signs. Table~\ref{tab:cols} lists the sixteen weights as
Standard Model fields. Throughout, a weight means a weight of a representation, as opposed to a Hamming weight.

\begin{table}[ht]
\centering\small
\begin{tabular}{@{}clc@{\qquad}clc@{}}
\toprule
column & field & $q$ & column & field & $q$\\
\midrule
1 & $d^c_1$ & $(+,-,-,-,-)$ & 8 & $d_1$ & $(-,+,+,-,+)$\\
2 & $d^c_2$ & $(-,+,-,-,-)$ & 9 & $d_2$ & $(+,-,+,-,+)$\\
3 & $u^c_3$ & $(-,-,+,+,+)$ & 10 & $u_3$ & $(+,+,-,+,-)$\\
4 & $d^c_3$ & $(-,-,+,-,-)$ & 11 & $d_3$ & $(+,+,-,-,+)$\\
5 & $u^c_2$ & $(-,+,-,+,+)$ & 12 & $u_2$ & $(+,-,+,+,-)$\\
6 & $u^c_1$ & $(+,-,-,+,+)$ & 13 & $u_1$ & $(-,+,+,+,-)$\\
7 & $e^c$ & $(+,+,+,-,-)$ & 14 & $e$ & $(-,-,-,-,+)$\\
 & & & 15 & $\nu$ (seed) & $(-,-,-,+,-)$\\
\bottomrule
\end{tabular}
\caption{The sixteen weights of the $\mathbf{16}$ as Standard Model fields, numbered in the order used
in the section on doubling, so that columns $\ell$ and $\ell+7$ are Dirac partners and column $15$ is the
seed. Subscripts are colour
labels, and the removed field is $\nu^c=(+,+,+,+,+)$.}
\label{tab:cols}
\end{table}

\paragraph{Relation to previous work.} The $SO(10)$ structure of a generation and its anomaly
cancellation, triorthogonal codes and doubling, and, for $p=2$, the fifteen-qubit code itself are
known. To our knowledge, the identification of anomaly cancellation with triorthogonality, the
$[[15,1,3]]_p$ codes for odd $p$, and the minimality and uniqueness of the qutrit code have not
appeared before.

That one generation with a right-handed neutrino forms the spinor of
$SO(10)$~\cite{Georgi1975,FritzschMinkowski1975}, labelled by five bits~\cite{Baez:2009dj}, and that
its anomalies cancel because $SO(10)$ has no cubic Casimir~\cite{Slansky1981}, is standard, as is the
triangle anomaly itself~\cite{Adler1969,BellJackiw1969,GrossJackiw1972}. General solutions of the
anomaly equations have been found for a single $U(1)$~\cite{CDF2019,AGT2020,GripaiosNguyen2026a}
and for two $U(1)$ factors~\cite{GripaiosNguyen2026b}, and the equations have also been solved for
extensions of the Standard Model gauge group~\cite{Allanach:2019uuu,AGT2020b}. Anomaly conditions for discrete gauge symmetries (which constrain the $\mathbb Z_p$ charges that label our codewords) go back
to~\cite{IbanezRoss1991,BanksDine1992}, with a more recent treatment
in~\cite{Hsieh:2018ifc}. We are not aware of earlier work connecting anomaly cancellation to
quantum codes.

On the coding side, triorthogonal codes were introduced for magic state
distillation~\cite{MSD,Bravyi_2012}, building on~\cite{knill2004faulttolerant}. The fifteen-qubit
code is a punctured Reed--Muller code~\cite{MSD}, and its relation to doubling is described
in~\cite{BravyiCross2015,Betsumiya_2012}. Short qubit triorthogonal codes have been constructed by
doubling~\cite{Jain_2025,Sullivan2024} and by other searches~\cite{Baldelli2026}, and small qubit
triorthogonal codes have been classified~\cite{Nezami_2022,WillsJainSingh2026}. For qudits, the
non-Clifford gate and distillation were developed
in~\cite{HowardVala,ACB,CampbellAnwarBrowne,campbell2014enhanced}, triorthogonal and related codes
in prime dimension in~\cite{Krishna_2019,PrakashCyclic,SahaNotes,Saha:2025frb,Sharma:2024zbn}, and
qutrit distillation schemes in~\cite{Prakash:2025azi,Prakash:2024ghq,PrakashCatalogue}.

Three other connections between codes and field theory are close in spirit to our construction. Gauss's
law in a lattice gauge theory is used for error detection or correction, with the gauge-invariant
states forming the code space~\cite{Stryker:2018efp,Rajput:2021trn,Spagnoli:2024mib}. Bao \emph{et al.}\ observed that superselection
sectors supply the orthogonality part of the Knill--Laflamme condition, protecting against bit flips
but not phase flips, and used the approximate isospin superselection of protons and neutrons in
low-energy QCD to define a code, concatenated with a repetition code to protect against phase
errors~\cite{Bao:2023lrb}. Gauge symmetry and charge conservation enter our construction too, since
for $p\ge5$ the stabilizers include the Cartan generators of the Standard Model gauge group, but play
a different role.  Here the qudits are labelled by fermion species,
the stabilizers are charge assignments rather than operators acting on states, and the property
that makes the spectrum a code is the exact cancellation of anomalies rather than gauge invariance
or an approximate superselection rule. Finally, real self-dual quantum stabilizer codes define
even self-dual Lorentzian lattices, and hence Narain conformal field theories, through a version of
Construction A~\cite{ConwaySloane,DymarskyShapere2021}. That correspondence relates codes and
lattices through a quadratic form. In this sense triorthogonal lattices are a cubic analogue, with
the anomaly in place of the inner product.

\section*{Part B: Proofs of the main-text results}

\subsection{Triorthogonal lattices and their reductions}

A lattice $\Lambda\subseteq\Z^n$ is \emph{triorthogonal}, for nonzero integer coefficients $c_\ell$, if the
cubic form $C(x)=\sum_\ell c_\ell x_\ell^3$ vanishes on it. Over $\Q$ a cubic form determines its trilinear form,
so this is the same as the vanishing of $\tau_c(x,y,z)=\sum_\ell c_\ell x_\ell y_\ell z_\ell$ on $\Lambda$. A
triorthogonal lattice reduces to a triorthogonal space modulo every prime that divides none of the
$c_\ell$, and linear algebra over $\Q$ controls the resulting codes for all but finitely many primes.

\begin{proposition}\label{prop:reduction}
Let $\Lambda$ be triorthogonal, with columns $h_\ell$, and let
$d_\Q=\min\{|J|:0\notin J,\ h_0\in\Span_\Q\{h_\ell :\ell\in J\}\}$. For all but finitely many odd primes $p$,
the reduction modulo $p$ punctured at coordinate $0$ is a triorthogonal $[[n-1,1,d_\Q]]_p$ code.
\end{proposition}

\begin{proof}
The cubic identity holds modulo every $p$. A logical $X$, a vector of the reduction with $x_0\neq0$,
exists unless $p$ divides every entry of $h_0$, and the logical exponent $\lambda=-c_0$ is nonzero
unless $p$ divides $c_0$. For the distance, take a relation over $\Q$ of minimal support that expresses
$h_0$ through $d_\Q$ other columns, scaled to a primitive integer vector. Unless $p$ divides one of its
finitely many coefficients, it survives modulo $p$ with the same support, so the distance is at most
$d_\Q$.

For the converse we use a rank argument, which recurs below. Let $J$ be a set of columns with
$h_0\notin\Span_\Q\{h_\ell:\ell\in J\}$, and let $\rho$ be the rank over $\Q$ of the matrix $[h_J]$
with these columns. Adding $h_0$ raises the rank to $\rho+1$, so the matrix $[h_0\,|\,h_J]$ has a
nonzero minor of size $\rho+1$. If $p$ does not divide that minor, $[h_0\,|\,h_J]$ still has rank
$\rho+1$ modulo $p$, while $[h_J]$ has rank at most $\rho$, so $h_0$ is not in the span of the columns
in $J$ modulo $p$. Applying this to the finitely many sets $J$ with $|J|<d_\Q$ excludes finitely many
primes, and for every other prime the distance is at least $d_\Q$.
\end{proof}

\subsection{The fifteen-qudit code}

Table~\ref{tab:H} gives a generator matrix of the code of Theorem~\ref{thm:sm}. The columns are the
fifteen fields, ordered as the ten weights of the $\mathbf{10}$ by the positions $ij$ of their two
minus signs and then the five weights of the $\bar{\mathbf 5}$ by the position $k$ of their plus sign.
The first row is the logical row $q_1$, and the other four are the Cartan generators $(q_i-q_{i+1})/2$ associated with the simple roots of
$\mathfrak{su}(5)$, the $X$-stabilizers. If $\nu^c$ is restored as the column $(1,0,0,0,0)$, the
cubic form of the integer matrix vanishes. The rows span a triorthogonal lattice,
which can be reduced modulo any odd prime.\footnote{It contains $\Lambda$ with index $2^4$,
because the stabilizer rows correspond to half-integral $t$, so the two lattices have the same
reduction modulo every odd prime.}

\begin{table}[h]
\centering\small
\setlength{\tabcolsep}{3.2pt}
\begin{tabular}{@{}l|rrrrrrrrrr|rrrrr@{}}
\toprule
 & $u^c_3$ & $u^c_2$ & $d_1$ & $u_1$ & $u^c_1$ & $d_2$ & $u_2$ & $d_3$ & $u_3$ & $e^c$ & $d^c_1$ & $d^c_2$ & $d^c_3$ & $\nu$ & $e$\\
 & $12$ & $13$ & $14$ & $15$ & $23$ & $24$ & $25$ & $34$ & $35$ & $45$ & $1$ & $2$ & $3$ & $4$ & $5$\\
\midrule
$q_1$ & $-1$&$-1$&$-1$&$-1$&$1$&$1$&$1$&$1$&$1$&$1$ & $1$&$-1$&$-1$&$-1$&$-1$\\
$(q_1-q_2)/2$ & $0$&$-1$&$-1$&$-1$&$1$&$1$&$1$&$0$&$0$&$0$ & $1$&$-1$&$0$&$0$&$0$\\
$(q_2-q_3)/2$ & $-1$&$1$&$0$&$0$&$0$&$-1$&$-1$&$1$&$1$&$0$ & $0$&$1$&$-1$&$0$&$0$\\
$(q_3-q_4)/2$ & $0$&$-1$&$1$&$0$&$-1$&$1$&$0$&$0$&$-1$&$1$ & $0$&$0$&$1$&$-1$&$0$\\
$(q_4-q_5)/2$ & $0$&$0$&$-1$&$1$&$0$&$-1$&$1$&$-1$&$1$&$0$ & $0$&$0$&$0$&$1$&$-1$\\
\bottomrule
\end{tabular}
\caption{A generator matrix of the fifteen-qudit code, with columns labelled by fields and, below, by
the positions of the minus signs ($\mathbf{10}$) or of the plus sign ($\bar{\mathbf 5}$) in the
weight. The first row is the logical row, and the other four are the $X$-stabilizers.}
\label{tab:H}
\end{table}

\paragraph{Qubits.} For $p=2$ we use the sign bits $b_i=(\one-q_{\cdot,i})/2$, $i=1,\dots,5$, of the
sixteen weights, together with $\one$. A binary code is triply even, that is, all its Hamming weights
are divisible by $8$, if it has generators whose weights are divisible by $8$, whose pairwise overlaps
are divisible by $4$ and whose triple overlaps are divisible by $2$~\cite{Bravyi_2012,Betsumiya_2012}.
Overlaps involving $\one$ reduce to overlaps without it, because $\one\wedge v=v$: thus
$|\one\wedge b_i|=|b_i|$ and $|\one\wedge b_i\wedge b_j|=|b_i\wedge b_j|$. Each reduced
quantity is already required to satisfy a stronger congruence (modulo $8$ rather than $4$, and
modulo $4$ rather than $2$). It therefore suffices to check $|\one|=n$ and, for distinct
$i,j,k$, the numbers $|b_i|$, $|b_i\wedge b_j|$ and $|b_i\wedge b_j\wedge b_k|$. 

Write $S_i=\sum_\ell q_{\ell,i}$,
$S_{ij}=\sum_\ell q_{\ell,i}q_{\ell,j}$ and $S_{ijk}=\sum_\ell q_{\ell,i}q_{\ell,j}q_{\ell,k}$. Since
$b_{\ell,i}=(1-q_{\ell,i})/2$, inclusion--exclusion gives
\begin{gather*}
|b_i|=\tfrac12(n-S_i),\qquad |b_i\wedge b_j|=\tfrac14(n-S_i-S_j+S_{ij}),\\
|b_i\wedge b_j\wedge b_k|=\tfrac18(n-S_i-S_j-S_k+S_{ij}+S_{ik}+S_{jk}-S_{ijk}).
\end{gather*}
For the generation $n=16$, and $S_i=S_{ijk}=0$ by anomaly cancellation, as in the main text, while
$S_{ij}=0$ for $i\neq j$. The three numbers are therefore $8$, $4$ and $2$, and the code is triply
even. The linear and quadratic sums are mixed anomalies with fermion number $F$, which assigns charge
$1$ to every field, and the cubic sum is the gauge anomaly. The same formulas show that, for any
spectrum of left-handed fermions with sign charges, the sign bits give a triply even code whenever, in
addition to the cubic anomalies, $n=\Tr F^3$ and $S_{ij}=\Tr(FQ_iQ_j)$ vanish modulo $16$; this is the
sense in which qubits need more than gauge anomaly cancellation.

The span of $\one$ and the $b_i$ is $RM(1,4)$ on the sixteen points of $\F_2^4$. Every $b_i$ vanishes
at the coordinate of $\nu^c$, because $q(\nu^c)=\one$, so a codeword $a\one+s$, with $s$ in the span of
the $b_i$, takes the value $a$ there. Its weight is divisible by $8$, so its weight on the other fifteen
coordinates is $\equiv-a\pmod8$, and $T^{\otimes15}$ multiplies every term of $\ket{\bar a}$ by
$e^{-i\pi a/4}$. Deleting the coordinate of $\nu^c$ thus gives the fifteen-qubit code~\cite{MSD}, on
which $T^{\otimes15}$ acts as $T^\dagger$ on the encoded qubit.

\subsection{Doubling and Dirac pairing}

Doubling builds a triorthogonal code from two ingredients: a $P$-code and a seed $\mathcal T$.

A $P$-code $(\mathcal S,\mu)$ is a subspace $\mathcal S\subseteq\F_p^{n_D}$ with nonzero
coefficients $\mu_\ell$ such that
\[
\sum_\ell\mu_\ell s_\ell=0\quad\text{and}\quad\sum_\ell\mu_\ell s_\ell s'_\ell=0\quad
\text{for all }s,s'\in\mathcal S,\qquad N=\sum_\ell\mu_\ell\neq0 .
\]
Its stabilizers are the vectors of $\mathcal S$, and its logical $X$ is $X(\one)$. A logical error of
$\mathcal S$ is a vector $v\in\mathcal S^\perp$ with $\sum_\ell v_\ell\neq0$, and the distance
$d_{\mathcal S}$ is the smallest weight of one.

The seed is a triorthogonal code $\mathcal T$ of length $n_{\mathcal T}$, with its own
coefficients and cubic form $\tau_{\mathcal T}$, logical row $y$ and stabilizer space $\mathcal T_0$.
Thus $\tau_{\mathcal T}$ vanishes on every triple of vectors of $\F_py+\mathcal T_0$ that contains a
vector of $\mathcal T_0$, and $\lambda_{\mathcal T}=\tau_{\mathcal T}(y,y,y)\neq0$ is its logical
exponent. A logical error of $\mathcal T$ is a vector $w\in\mathcal T_0^\perp$ with $y\cdot w\neq0$,
and $d_{\mathcal T}$ is the smallest weight of one.

The doubled code has three blocks of columns, of widths $n_D$, $n_D$ and $n_{\mathcal T}$: a first copy
of $\mathcal S$, a second copy of $\mathcal S$, with columns of one copy paired with columns
of the other, and the seed. In our convention, both copies carry the same coefficients $\mu$, and the seed has its original coefficients. The signs are as per the generator matrix below, which depends on a parameter $\kappa$,
\[
\begin{array}{lccc}
 & \text{first copy} & \text{second copy} & \text{seed}\\
\text{logical row } L & -\kappa\one & \kappa\one & y\\
\text{paired stabilizers} & s & -s & 0\\
\text{linking row } K & 0 & \kappa\one & y\\
\text{seed stabilizers} & 0 & 0 & t
\end{array}
\]
with $s$ running over a basis of $\mathcal S$ and $t$ over a basis of $\mathcal T_0$. The rows $L$ and
$K$ agree on the second copy and the seed, and differ only on the first copy, where $K$ vanishes.

\begin{proposition}\label{prop:doubling}
If $N\kappa^3+\lambda_{\mathcal T}=0$, the doubled matrix, with coefficients $\mu$ on both copies, is
triorthogonal with logical exponent $\lambda_{\mathcal T}$, and its distance is
$\min(d_{\mathcal S},d_{\mathcal T}+2)$.
\end{proposition}

\begin{proof}
\emph{Triorthogonality.} A general vector of the row space is
\[
u=\alpha L+\gamma K+(s,-s,0)+(0,0,t)
=\big(-\alpha\kappa\one+s,\ \beta\kappa\one-s,\ \beta y+t\big),\qquad \beta=\alpha+\gamma,
\]
with $s\in\mathcal S$ and $t\in\mathcal T_0$. Its stabilizers are the vectors with $\alpha=0$, so we
must show that $\tau(u,u',u'')=\lambda_{\mathcal T}\,\alpha\alpha'\alpha''$ for any three such vectors
$u$, $u'$ and $u''$. The form $\tau$ is a sum over columns, so it splits into one term for each block.
On the first copy, expand the product of the three entries. The terms with one or two factors from
$\mathcal S$ vanish by the $P$-code conditions, leaving
\[
-\alpha\alpha'\alpha''N\kappa^3+\sum_\ell\mu_\ell s_\ell s'_\ell s''_\ell .
\]
On the second copy the same expansion gives
\[
\beta\beta'\beta''N\kappa^3-\sum_\ell\mu_\ell s_\ell s'_\ell s''_\ell ,
\]
because $s$ enters with the opposite sign, and the last term cancels the one from the first copy. On the
seed, the triorthogonality of $\mathcal T$ removes every term that contains $t$, $t'$ or $t''$,
leaving $\beta\beta'\beta''\lambda_{\mathcal T}$. Altogether,
\[
\tau(u,u',u'')=-\alpha\alpha'\alpha''N\kappa^3+\beta\beta'\beta''\big(N\kappa^3+\lambda_{\mathcal T}\big).
\]
When $N\kappa^3+\lambda_{\mathcal T}=0$, the second term vanishes and the first is
$\lambda_{\mathcal T}\,\alpha\alpha'\alpha''$, as required.

\emph{Distance.} Write a $Z$-error as $z=(z_1,z_2,z_3)$, on the first copy, the second copy and the
seed. It commutes with the paired stabilizers, the seed stabilizers and $K$ when, respectively,
\[
\text{(i)}\ \ z_1-z_2\in\mathcal S^\perp,\qquad
\text{(ii)}\ \ z_3\in\mathcal T_0^\perp,\qquad
\text{(iii)}\ \ \kappa\textstyle\sum z_2+y\cdot z_3=0 .
\]
Given (iii), $L\cdot z=-\kappa\sum z_1+\kappa\sum z_2+y\cdot z_3=-\kappa\sum z_1$, and $\kappa\neq0$
because $N\kappa^3=-\lambda_{\mathcal T}\neq0$. So $z$ is a logical error exactly when (i)--(iii) hold
and $\sum z_1\neq0$. There are two cases.
\begin{itemize}
\item If $y\cdot z_3=0$, then $\sum z_2=0$ by (iii), so $v=z_1-z_2$ lies in $\mathcal S^\perp$ with
$\sum v=\sum z_1\neq0$ -- a logical error of $\mathcal S$. Hence
$\wt(z)\ge\wt(z_1)+\wt(z_2)\ge\wt(v)\ge d_{\mathcal S}$.
\item If $y\cdot z_3\neq0$, then by (ii) $z_3$ is a logical error of $\mathcal T$, so
$\wt(z_3)\ge d_{\mathcal T}$. Moreover $z_1\neq0$, since $\sum z_1\neq0$, and $z_2\neq0$, since
$\kappa\sum z_2=-y\cdot z_3\neq0$. Hence $\wt(z)\ge d_{\mathcal T}+2$.
\end{itemize}
So the distance is at least $\min(d_{\mathcal S},d_{\mathcal T}+2)$, and both bounds are attained. If
$v$ is a minimum-weight logical error of $\mathcal S$, then $z=(v,0,0)$ satisfies (i)--(iii) and has
$\sum z_1\neq0$. If $w$ is a minimum-weight logical error of $\mathcal T$, let $z_1=z_2$ equal
$-(y\cdot w)/\kappa$ on one column and zero elsewhere. Then $z=(z_1,z_2,w)$ satisfies (i), since
$z_1-z_2=0$, as well as (ii) and (iii), and has $\sum z_1=-(y\cdot w)/\kappa\neq0$.
\end{proof}

\paragraph{Proof of Theorem~\ref{thm:dirac}.} The proof is a change of basis. We build two copies of
a seven-qudit $P$-code and a one-qudit seed from pieces of the generation, check the conditions of
Proposition~\ref{prop:doubling}, and show that the doubled matrix generates the same lattice as
Eq.~\eqref{eq:H}. Every identity below holds over $\Z$, and so modulo every odd $p$.

The generation splits into two Pati--Salam multiplets~\cite{PatiSalam1974}. The right-handed sector
$\mathcal R$, the fields $f^c$ and $\nu^c$, is a $(\bar{\mathbf 4},\mathbf 1,\mathbf 2)$ of
$SU(4)\times SU(2)_L\times SU(2)_R$. On it $T_{3L}=0$ and $q_4=q_5=2T_{3R}$. Its image under $\sigma$,
the Dirac partners $f$ and $\nu$, is a $(\mathbf 4,\mathbf 2,\mathbf 1)$. On it $T_{3R}=0$ and
$q_4=-q_5=2T_{3L}$. We order the fields as in Table~\ref{tab:cols}, so that columns $\ell$ and
$\ell+7$ hold $f^c$ and $f=\sigma(f^c)$, and column $15$ holds $\nu$.

\emph{The right-handed code.} Remove $\nu^c$ from $\mathcal R$, and give each of the seven remaining
fields the coefficient $\mu=2T_{3R}=q_4$. The charges under which $\nu^c$ is neutral are spanned by
$(q_5-q_i)/2$, $i=1,2,3$; these are the stabilizers. The logical row is $2T_{3R}$, which is $1$ at
$\nu^c$. Proposition~\ref{prop:doubling} takes the logical row to be $\one$, so we multiply each
column by its value of $q_4$. Physically, this replaces $d^c$ and $e^c$ by their conjugates. The
quadratic form is unchanged, since $q_4^2=1$. The stabilizers become
$s_i=q_4(q_5-q_i)/2=(1-q_4q_i)/2$, which is $1$ exactly when colour coordinate $i$ of the field
differs from its weak coordinates. In the order of Table~\ref{tab:cols} the code is
\begin{equation}\label{eq:steane}
\setlength{\arraycolsep}{4pt}
\begin{array}{r|ccccccc}
 & d^c_1 & d^c_2 & u^c_3 & d^c_3 & u^c_2 & u^c_1 & e^c\\ \hline
\text{logical} & 1&1&1&1&1&1&1\\
s_1 & 1&0&1&0&1&0&1\\
s_2 & 0&1&1&0&0&1&1\\
s_3 & 0&0&0&1&1&1&1\\ \hline
\mu=2T_{3R} & \bar1&\bar1&1&\bar1&1&1&\bar1
\end{array}
\end{equation}
The columns of the $s_i$ are the binary numbers $1,\dots,7$. The stabilizers therefore span the
$[7,3,4]$ simplex code, and for $p=2$ this is the Steane code. The $P$-code conditions can be read
off directly. Each $s_i$ is supported on four columns and each $s_i*s_j$, $i\neq j$, on two, and in
each case half of them have $\mu=+1$ and half $\mu=-1$. Moreover $N=\sum\mu=-1=-2T_{3R}(\nu^c)$,
the anomaly that $\nu^c$ would cancel.

The distance is $3$ for every $p$. The columns of the $s_i$ are distinct nonzero $0/1$ vectors, so no
two are parallel and no error of weight one or two commutes with the stabilizers. Columns $1$, $2$
and $3$ give the logical error $(1,1,\bar1,0,0,0,0)$, which is orthogonal to every $s_i$ and has sum
$1$. Undoing the rescaling, it is the relation $\nu^c=-d^c_1-d^c_2-u^c_3$ of Table~\ref{tab:ops}. The
right-handed sector is thus a $[[7,1,3]]^P_p$ code.

\emph{The Dirac partners.} The map $\sigma$ takes $\mathcal R$ to the left-handed doublets and $\nu^c$
to $\nu$. It preserves $q_4$ and negates $q_i-q_5$, so it takes $2T_{3R}$ to $2T_{3L}$ and each
stabilizer to its negative. After the same rescaling, the column of $f$ therefore has the same
coefficient as that of $f^c$, since $2T_{3L}(f)=2T_{3R}(f^c)$, and stabilizer entries $-s_i$.
Without $\nu$, which becomes the seed, the doublets form a second copy of the same code, with every
stabilizer negated on paired columns, which is exactly the pairing that doubling requires. Physically,
each $f$ has the colour and electric charge opposite to those of $f^c$, and the stabilizers span,
over $\Q$, the Cartan algebra of $SU(3)_c\times U(1)_{\rm em}$. Their differences give $q_1-q_2$ and
$q_2-q_3$, and their sum gives $Q_{\rm em}=T_{3L}+Y=-\tfrac13\sum_i(q_5-q_i)/2$. On these charges the
generation is vectorlike.

\emph{The seed.} The neutrino $\nu$ is the one field whose Dirac partner, $\nu^c$, is not among the
qudits. On its own it is the trivial code: one qudit, coefficient $1$, no stabilizers, and logical
row $y=(1)=2T_{3L}(\nu)$. Thus $\lambda_{\mathcal T}=1$ and $d_{\mathcal T}=1$.

\emph{Doubling.} The linking row must vanish on the first copy and equal $\kappa\one$ on the second
copy and $y$ on the seed. Weak isospin $2T_{3L}=(q_4-q_5)/2$ does this with $\kappa=1$, because the
fields $f^c$ are weak singlets. Since paired columns carry equal coefficients, $N=\sum_f2T_{3L}(f)$,
and the condition of Proposition~\ref{prop:doubling} becomes
\[
N\kappa^3+\lambda_{\mathcal T}=\Tr_{\sigma(\mathcal R)}2T_{3L}=-1+1=0 .
\]
Weak isospin is traceless on the doublets. By Proposition~\ref{prop:doubling}, the doubled code is
triorthogonal, with logical exponent $1$ and distance $\min(3,1+2)=3$. With coefficients $\mu$ on both
copies and $1$ on the seed, its generator matrix is
\begin{equation}\label{eq:doubled}
\setlength{\arraycolsep}{2.5pt}
\begin{array}{r|ccccccc|ccccccc|c}
 & \multicolumn{7}{c|}{f^c} & \multicolumn{7}{c|}{f} & \nu\\ \hline
L & \bar1&\bar1&\bar1&\bar1&\bar1&\bar1&\bar1 & 1&1&1&1&1&1&1 & 1\\
s_1 & 1&0&1&0&1&0&1 & \bar1&0&\bar1&0&\bar1&0&\bar1 & 0\\
s_2 & 0&1&1&0&0&1&1 & 0&\bar1&\bar1&0&0&\bar1&\bar1 & 0\\
s_3 & 0&0&0&1&1&1&1 & 0&0&0&\bar1&\bar1&\bar1&\bar1 & 0\\
K & 0&0&0&0&0&0&0 & 1&1&1&1&1&1&1 & 1
\end{array}
\end{equation}
For $p=2$ this is the doubled Steane code~\cite{BravyiCross2015}.

\emph{Comparison with Eq.~\eqref{eq:H}.} Undo the rescaling by negating the columns of $d^c$, $e^c$,
$d$ and $e$. The coefficients all return to $1$, and the rows become charges of the generation:
\[
L=2(T_{3L}-T_{3R})=-q_5,\qquad \tilde s_i=\tfrac12(q_5-q_i),\qquad K=2T_{3L}=\tfrac12(q_4-q_5),
\]
where $\tilde s_i$ is the row $s_i$ with the rescaling undone. In terms of these,
\[
q_1=-L-2\tilde s_1,\quad \tfrac{q_1-q_2}2=\tilde s_2-\tilde s_1,\quad \tfrac{q_2-q_3}2=\tilde s_3-\tilde s_2,\quad
\tfrac{q_3-q_4}2=-\tilde s_3-K,\quad \tfrac{q_4-q_5}2=K .
\]
This change of basis is unimodular, so with the rescaling undone the rows of Eq.~\eqref{eq:doubled} and
of Eq.~\eqref{eq:H} generate the same lattice over $\Z$, with columns matched as in Tables~\ref{tab:cols}
and~\ref{tab:H}. The stabilizers, the rows that vanish at $\nu^c$, correspond to each other. The
logical row $q_1$ of Eq.~\eqref{eq:H} is $-L$ modulo stabilizers, which relabels the logical basis by
$a\to-a$. The transversal gate, logical $U$ in the doubled form, is therefore $U^\dagger$ in the basis
of Theorem~\ref{thm:sm}, since $(-a)^3=-a^3$. 

\paragraph{Integral doubling.} For qubits, doubling maps doubly even codes to triply even
ones~\cite{Betsumiya_2012}, so it preserves the integral refinement that makes the transversal gate
exact. For odd $p$ the same holds more simply. Call
$(\mathcal S,\mu)$ an \emph{integral $P$-lattice} if $\mathcal S\subseteq\Z^{n_D}$ is a lattice and
$\mu\in\Z^{n_D}$ has nonzero entries satisfying the $P$-code conditions over $\Z$, with $N\neq0$.

\begin{proposition}
Let $(\mathcal S,\mu)$ be an integral $P$-lattice, let $\mathcal T$ be an integral triorthogonal
lattice with logical exponent $\lambda_{\mathcal T}$, and let $\kappa\in\Z$ satisfy
$N\kappa^3+\lambda_{\mathcal T}=0$. Then the doubled matrix, with coefficients $\mu$ on both copies, is
an integral triorthogonal lattice with logical exponent $\lambda_{\mathcal T}$.
\end{proposition}

\begin{proof}
The triorthogonality computation in the proof of Proposition~\ref{prop:doubling} is an identity in
the entries, so it holds over $\Z$.
\end{proof}

With unit coefficients, the doubled lattice is the charge lattice of an anomaly-free spectrum, in which
the two copies are vectorlike pairs under the logical row and the paired stabilizers and chiral under
the linking row. Theorem~\ref{thm:dirac} is the case in which $\mathcal S$ is the right-handed sector,
$\mu=2T_{3R}$, $N=-1$, $\kappa=1$ and $\mathcal T$ is the trivial seed. Doubling can therefore be
iterated within anomaly-free spectra; every code obtained in this way has an exact qutrit gate
(Proposition~\ref{prop:dict}), and its reductions are triorthogonal codes for all but finitely many
primes, with the distance of the lattice over $\Q$.

There is one restriction. Over $\Z$, the condition $\sum_\ell\mu_\ell s_\ell^2=0$ forces $s=0$ when all
$\mu_\ell>0$, so a nonzero integral $P$-lattice has coefficients of both signs. In particular, a
$P$-code whose coefficients are all $1$, such as the ternary Golay code, never lifts with unit
coefficients. For unit coefficients and $\lambda_{\mathcal T}=\pm1$, the condition on $\kappa$ forces
$|N|=1$. An integral $P$-lattice of distance five with $|N|=1$ is the input that would double the
fifteen-qudit code to distance five.

\subsection{Minimality, uniqueness and symmetry}

Here $p=3$. We use the matrix form of triorthogonality~\cite{PrakashCyclic}. A generator matrix $H$
with logical rows $g_i$ and stabilizer rows spanning $A$ is triorthogonal if $\tau_c$ vanishes on
every triple of rows except a logical row taken three times, and $\tau_c(g_i,g_i,g_i)\neq0$. The
distance is $d=\min\{\wt(v):v\in A^\perp\setminus H^\perp\}$. For one logical qutrit this is
equivalent to the definition in the main text, and it does not impose self-orthogonality.

\begin{proposition}\label{prop:unique}
No triorthogonal qutrit code with $d\ge3$ has length at most $14$. Every triorthogonal
$[[15,1,3]]_3$ code is equivalent to the code of Theorem~\ref{thm:sm}, and there is no
triorthogonal $[[15,k,3]]_3$ code with $k\ge2$ and no $[[15,1,4]]_3$ code. The monomial automorphism
group of the code of Theorem~\ref{thm:sm} is $S_5\times\Z_2$. This strengthens Theorem~\ref{thm:unique}.
\end{proposition}

\paragraph{Reduction.} We first bring a shortest code with $d\ge3$ to a normal form. None of the
following steps lowers the distance or lengthens the code.
\begin{itemize}
\item \emph{One logical row.} If there are several logical rows, keep one, $g$. This leaves $A$
unchanged and only shrinks $H$, so the set $A^\perp\setminus H^\perp$ of logical errors shrinks and
$d$ cannot decrease.
\item \emph{No zero or parallel columns.} A zero column of $H$ contributes nothing to $\tau_c$ and can
be deleted. Two parallel columns, $h$ and $\beta h$ with coefficients $c$ and $c'$, contribute to
$\tau_c$ as a single column $h$ with coefficient $c+c'\beta^3$, and can be replaced by it, or deleted
if that coefficient vanishes. In each case a logical error of the shorter code, extended by zero, is a
logical error of the original code of the same weight, so $d$ cannot decrease. A shortest code with
$d\ge3$ therefore has no zero or parallel columns.
\item \emph{Unit coefficients.} Multiplying column $\ell$ by $c_\ell$ and setting its coefficient to
$1$ leaves $\tau_c$ unchanged, because $c_\ell^3=c_\ell$ in $\F_3$, and it does not change Hamming
weights.
\item \emph{Projectivity.} Now $d\ge3$ exactly when the columns of $A$ are nonzero and pairwise
non-parallel, that is, when $A$ is projective. A $Z$-error of weight one or two lies in $A^\perp$
exactly when the columns of $A$ in its support are linearly dependent, and it then lies outside
$H^\perp$, because the corresponding columns of $H$ are independent; it is a logical error.
\item \emph{Normal form.} Row reduction and a permutation of columns bring $A$ to the form
$(I_r\,|\,P)$, and subtracting stabilizers from $g$ makes it vanish on the first $r$ columns, so that
$g=(0\,|\,w)$, with $P\in\F_3^{r\times m}$ and $n=r+m$. Triorthogonality becomes polynomial conditions
on $P$ and $w$, one for each triple of rows.
\end{itemize}
Two bounds then limit $(r,m)$. The vectors $x_i*x_i$, with $x_i$ the rows of $A$ and $*$ the entrywise
product, lie in $H^\perp$, since $(x_i*x_i)\cdot h=\tau_c(x_i,x_i,h)=0$ for every row $h$ of $H$, and
they are independent, since on the identity block $x_i*x_i$ is nonzero only at position $i$. As
$H^\perp$ has dimension $n-r-1=m-1$, this gives $m\ge r+1$, a cubic analogue of the bound
$\dim C\le n/2$ for self-orthogonal codes. A projective $A$ has at most as many
columns as there are points in the projective space $PG(r-1,3)$, so $r+m\le\frac12(3^r-1)$. Together
these force $r\ge3$.

\paragraph{Search.} We enumerate canonical representatives, lexicographically minimal under
permutations of the columns of $\binom wP$, permutations of the rows of $P$, and $w\to-w$, by
backtracking over the rows of $P$. Every normal form is equivalent to a canonical one, so the search is
exhaustive. For $n\le14$ the bounds leave $19$ pairs $(r,m)$, and none has a solution, so no code with
$d\ge3$ has length at most $14$. For $n=15$ only $(r,m)=(4,11)$ has solutions, $36$ of them.

Every $[[15,1,3]]_3$ code appears among these solutions. It has no zero or parallel columns, since
removing them would give a shorter code with $d\ge3$, and the other reduction steps are equivalences.
Each of the $36$ solutions is equivalent to the code of Theorem~\ref{thm:sm}. For each, we find a
permutation of columns that maps its row space onto that of Table~\ref{tab:H}. Since all coefficients
are $1$, such a permutation preserves $\tau_c$, and so it also maps stabilizers to stabilizers, because
the stabilizer space is determined by $\tau_c$, as the set of row-space vectors $x$ with
$\tau_c(x,y,z)=0$ for all $y,z$ in the row space. All $36$ solutions have $d=3$, so there is no
$[[15,1,4]]_3$ code.

A $[[15,k,3]]_3$ code with $k\ge2$ would, by keeping one logical row, contain one of these codes,
together with a second logical row $g'$ outside the span of the stabilizers and $g$ that satisfies
$\tau_c(a,b,g')=\tau_c(a,g,g')=\tau_c(g,g,g')=0$ for all stabilizers $a,b$. These conditions are
linear in $g'$, and for each solution the vectors satisfying them are exactly the stabilizers, so no
such $g'$ exists. The number of partial matrices the
search visits for each $(r,m)$, and a check against brute-force enumeration for small $(r,m)$, are in
the repository.

\paragraph{Automorphisms.} Magma~\cite{Magma} gives a monomial automorphism group of order $240$,
consisting of the permutations of the five coordinates of $q$ together with an overall sign. These
permutations fix $\nu^c=\one$, and they form its whole stabilizer in the Weyl group of $SO(10)$. That
group has order $1920$ and acts transitively on the sixteen weights, so the stabilizer has order
$1920/16=120=|S_5|$. For $p=5$, $7$ and $11$ the group is
$S_5\times\F_p^\times$, so again no symmetries appear beyond the Weyl group and global scalars.

\subsection{Distillation performance}

Noisy copies of the magic state are twirled so that each suffers a random
$Z^k$, $k\neq0$, with total probability $\epsilon$, its infidelity, each nonzero power being equally
likely~\cite{ACB,CampbellAnwarBrowne}. Depolarizing noise of strength $\delta$, which replaces the state
by the maximally mixed state with probability $\delta$, gives $\epsilon=\frac{p-1}{p}\,\delta$, which
is $\delta/2$ for qubits and $2\delta/3$ for qutrits. We compare the two codes both at equal
$\epsilon$, the usual figure of merit for distillation, and at equal $\delta$, which holds the channel
fixed. The qutrit code is better in both, by a factor of about $5.6$ in the leading term at equal
$\epsilon$ and about $3$ at equal $\delta$.

A round of distillation is accepted when the $Z$-error lies in $A^\perp$ and fails when it lies
in $A^\perp\setminus H^\perp$. For the qubit code the weight-three failures are the $35$ supports of
weight-three codewords of the Hamming code, each with one error pattern, so
$\epsilon_{\rm out}\approx35\,\epsilon^3$. For the qutrit code they are the $25$ supports of Table~\ref{tab:ops}. On each support, $2$ of the $2^3=8$ patterns of $Z$ and $Z^2$ are logical (the relation
and its double), and the other $6$ are detected, so $\epsilon_{\rm out}\approx50(\epsilon/2)^3=6.25\,\epsilon^3$.
In terms of $\delta$, with $\delta_{\rm out}=\frac{p}{p-1}\epsilon_{\rm out}$, these become
$8.75\,\delta^3$ and $2.78\,\delta^3$. Although a qutrit has more error types than a qubit, a smaller
fraction of them are logical.

Table~\ref{tab:distill} gives the exact acceptance
probability, output error and threshold of one round of each code, computed from the weight
enumerators of the accepted errors. These begin $1+50z^3+260z^4+\cdots$ for the qutrit code, of which
the $50$ weight-three terms are all logical, and $1+35z^3+105z^4+\cdots$ for the qubit code. At equal
infidelity the two codes accept with the same probability to leading order, $1-15\epsilon$.

\begin{table}[h]
\centering\small
\begin{tabular}{@{}lcc@{}}
\toprule
 & $[[15,1,3]]_3$ & $[[15,1,3]]_2$\\
\midrule
weight-three logical supports & $25$ & $35$\\
logical patterns per support & $2$ of $8$ & $1$ of $1$\\
leading term, equal infidelity & $6.25\,\epsilon^3$ & $35\,\epsilon^3$\\
leading term, equal $\delta$ & $2.78\,\delta^3$ & $8.75\,\delta^3$\\
threshold $\epsilon^*$ & $0.230$ & $0.141$\\
threshold $\delta^*$ & $0.345$ & $0.283$\\
acceptance at $\epsilon=0.01$ & $0.860$ & $0.860$\\
acceptance at $\delta=0.01$ & $0.905$ & $0.928$\\
$\epsilon$ after two rounds from $\epsilon=10^{-2}$ & $1.7\times10^{-15}$ & $1.6\times10^{-12}$\\
\bottomrule
\end{tabular}
\caption{Single-code distillation with the two fifteen-qudit codes, computed exactly.}
\label{tab:distill}
\end{table}

Both codes have yield exponent $\log15/\log3\approx2.46$, so the advantage is a
constant factor. We conclude with an obvious caveat to this comparison: across dimensions, noisy qutrit and qubit magic states need not cost the same to prepare and
the distilled gates differ.

\section*{Part C: Physical interpretation and further results}

\subsection{The minimal logical errors as four-fermion operators}

By Eq.~\eqref{eq:distance}, a minimum-weight $Z$-logical of the code of Theorem~\ref{thm:sm} is a
linear relation $q(\nu^c)=\sum_fs_f\,q(f)$ among weights, where $q(f)$ is the weight of the field $f$
(Table~\ref{tab:cols}). Each such relation is a statement of charge conservation. The product of
$\nu^{c\dagger}$ with the fields $f$ for which $s_f=+1$ and the conjugates $f^\dagger$ for which
$s_f=-1$ has zero charge under every Cartan generator of $SO(10)$ (Part A), and in particular under
those of the Standard Model. We call such a product a Cartan-neutral four-fermion operator. In general
it is not, by itself, invariant under the full Standard Model gauge group, which also mixes colour and
isospin components.

Since $d=3$ (Theorem~\ref{thm:sm}), every such relation involves exactly three fields besides $\nu^c$,
and by direct computation all its coefficients are $s_f=\pm1$. There are $25$, one for each weight-three support, listed in
Table~\ref{tab:ops}. 

Each relation also has the charge structure of a tree-level exchange, which the weights make
explicit. In $SO(10)$, two fermions in the $\mathbf{16}$ can couple to a scalar in the $\mathbf{10}$,
the smallest representation in $\mathbf{16}\otimes\mathbf{16}=\mathbf{10}\oplus\mathbf{120}\oplus
\mathbf{126}$, which is how Yukawa couplings arise in $SO(10)$ models. Such a coupling conserves charge
only if the two fermions' charge vectors add up to a weight of the $\mathbf{10}$. In the normalization
used here, these weights are $\pm2\hat e_i$, where $\hat e_i$ is the
$i$-th unit vector, for example $\hat e_1=(1,0,0,0,0)$. Two sign vectors therefore qualify exactly when they agree in one
coordinate and are opposite in the other four. Exchanging the scalar between two such couplings gives
a four-fermion operator whose fields split into two pairs of this kind. Every relation in
Table~\ref{tab:ops} splits in this way, with $\nu^c$ in one of the pairs; for conjugated fields the
charge vector is $-q$. For example, the right-handed current of Fig.~\ref{fig:code} splits into
$e^c=(+,+,+,-,-)$ with $u^c_1=(+,-,-,+,+)$, which agree only in the first coordinate, and
$\nu^{c\dagger}$ with $d^{c\dagger}_1$.

The shared coordinate says which component of the $\mathbf{10}$ is exchanged. Under $SU(5)$ the
$\mathbf{10}$ is a $\mathbf 5\oplus\bar{\mathbf 5}$. A weak coordinate ($i=4,5$) corresponds to a Higgs
doublet, and a colour coordinate ($i=1,2,3$) to a colour triplet. The products of Dirac masses and the
mixed and scalar currents can arise from Higgs-doublet exchange, through the same Yukawa couplings that
give fermions their masses. The right-handed currents and the nine baryon-number-violating relations
require colour-triplet exchange, which mediates proton decay in $SO(10)$ models. Similarly, a pair
$\psi\psi'^\dagger$ can be joined by an $SO(10)$ gauge boson, in the adjoint
$\mathbf{45}\subset\mathbf{16}\otimes\bar{\mathbf{16}}$, when $q(\psi)-q(\psi')$ is a root
$\pm2\hat e_i\pm2\hat e_j$, that is, when the two weights differ in exactly two coordinates. The fifteen relations
marked $g$, those with coefficients of both signs, split into two such pairs, so they also have the
charge structure of gauge-boson exchange.

\begin{table}[h]
\centering\small
\begin{tabular}{@{}llcc@{}}
\toprule
type & relation & number & gauge\\
\midrule
Dirac bilinears $(\nu\nu^c)(uu^c)^\dagger$ & $\nu^c=u_i+u^c_i-\nu$ & 3 & $g$\\
Dirac bilinears $(\nu\nu^c)(dd^c)$ & $\nu^c=-d_i-d^c_i-\nu$ & 3 & \\
Dirac bilinears $(\nu\nu^c)(ee^c)$ & $\nu^c=-e-e^c-\nu$ & 1 & \\
right-handed current & $\nu^c=e^c+u^c_i-d^c_i$ & 3 & $g$\\
mixed current & $\nu^c=d_i+u^c_i-e$ & 3 & $g$\\
$\Delta B=\Delta L$ & $\nu^c=d_i+u_j-d^c_k$ & 6 & $g$\\
$\Delta B=\Delta L$ & $\nu^c=-d^c_i-d^c_j-u^c_k$ & 3 & \\
scalar charged current & $\nu^c=-u_i-d^c_i-e$ & 3 & \\
\midrule
total & & 25 & 15\\
\bottomrule
\end{tabular}
\caption{The $25$ minimum-weight logicals as four-fermion relations. Subscripts are colours. Equal
subscripts denote the same colour, and $i,j,k$ are distinct, so $\nu^c=d_i+u_j-d^c_k$ occurs for the
six orderings of three colours. All relations can arise from exchange of a $\mathbf{10}$, and those
marked $g$ also arise from exchange of a gauge boson.}
\label{tab:ops}
\end{table}

\subsection{Which fermions are removed}

Because the Weyl group of $SO(10)$ acts transitively on the sixteen weights, removing any one of them
gives an equivalent code. We single out $\nu^c$ by gauge symmetry, demanding that the $X$-stabilizers
include the Cartan generators of the Standard Model gauge group $SU(3)_c\times SU(2)_L\times U(1)_Y$.
Since the $X$-stabilizers are the symmetries under which the removed fermion is neutral, the removed
fermion must be neutral under all of these generators. As integral generators we take $q_1-q_2$,
$q_2-q_3$, $(q_4-q_5)/2$ and $6Y=3n_w-2n_c=q_1+q_2+q_3-\tfrac32(q_4+q_5)$, where $n_c$ and $n_w$ are
the numbers of occupied colour and weak oscillators. Neutrality under the colour generators requires
$q_1=q_2=q_3$, so $n_c\in\{0,3\}$, and neutrality under the weak generator requires $q_4=q_5$, so
$n_w\in\{0,2\}$. Since every field of the $\mathbf{16}$ has $n_c+n_w$ even, $n_c=0$, and the removed
field is either $\nu^c$, with $n_w=0$ and $6Y=0$, or the positron $e^c$, with $n_w=2$ and $6Y=6$. Over
$\Z$, or modulo any prime $p\ge5$, only $\nu^c$ is neutral, and this is the field we remove. Modulo $3$
the positron is neutral too. Either way, the removed field is the singlet of an $\mathfrak{su}(5)$ whose
Cartan generators are the stabilizers and whose $\mathbf{10}$ and $\bar{\mathbf 5}$ are the two orbits
of the remaining fields. The singlet is $\nu^c$ in the Georgi--Glashow
embedding~\cite{GeorgiGlashow1974}, and $e^c$ in flipped $SU(5)$~\cite{Barr1982,Derendinger1984},
where hypercharge combines an $\mathfrak{su}(5)$ generator with $U(1)_\chi$.

We can also remove several fermions with independent weights, to obtain higher-rate codes, at the cost of reduced distance. Removing $k$ of them gives a
triorthogonal code with $k$ logical qudits, whose stabilizers are the symmetries under which every
removed fermion is neutral. Triorthogonality of the full lattice makes $U^{\otimes(16-k)}$ act as
$U^\dagger$ on each logical qudit, exactly for $p=3$ by the argument of
Proposition~\ref{prop:dict}, and $d=1$ exactly when some remaining fermion completes a neutral
combination with removed ones. Removing any two fermions gives a $[[14,2,2]]_p$ code for every $p$;
for qubits it is the first member of the Bravyi--Haah family~\cite{Bravyi_2012}. For odd $p$,
removing three keeps $d\ge2$ exactly when they belong to one $SU(4)$ quartet of Pati and
Salam~\cite{PatiSalam1974}, in which lepton number is the fourth colour, such as
$\{\nu^c,u^c_1,u^c_2,u^c_3\}$, or to a Weyl image of one, and gives a $[[13,3,2]]_p$ code. Any other
triple lies in a four-fermion relation, a Weyl image of one in Table~\ref{tab:ops}, whose fourth field
gives $d=1$. Removing a whole quartet also gives $d=1$, because $\nu^c-u^c_1-u^c_2-u^c_3=2e^c$. For
qubits any removal of three gives $d=1$, and for qutrits any removal of four gives $d=1$. (For $p \geq 5$, it is possible to find an additional removal to obtain a $[[12,4,2]]_p$ code, but we do not offer a Standard Model interpretation for this.)

\subsection{Other representations}

We now ask which representations other than the $\mathbf{16}$ give useful codes. The first
requirement is anomaly freedom. Among the compact simple Lie algebras only $\mathfrak{su}(m)$ with
$m\ge3$ has a cubic Casimir~\cite{Slansky1981}, so every representation of every other one is anomaly
free.

The second requirement is chirality. Real and pseudoreal representations contain each weight together
with its negative, so every field has a vectorlike partner and $d=1$. Several identical generations
fail for a similar reason, since every field has an identical copy. Among the algebras without a cubic
Casimir, only $\mathfrak{so}(4k+2)$ with $k\ge2$ and $\mathfrak e_6$ have complex representations. The
smallest are the spinors, starting with the $\mathbf{16}$ of $\mathfrak{so}(10)$, and the $\mathbf{27}$
of $\mathfrak e_6$. 

The $\mathbf{27}$ gives $d=2$. Under $SO(10)\times U(1)\subset E_6$ it decomposes as
$\mathbf{16}_1\oplus\mathbf{10}_{-2}\oplus\mathbf 1_4$, so in the normalization of the main text its
weights are $(q,1)$ for the sixteen weights $q$ of the $\mathbf{16}$, $(\pm2\hat e_i,-2)$ and $(0,4)$.
No weight is a multiple of another, even modulo an odd prime. The $\mathbf{27}$ carries an invariant
cubic form of its own, distinct from a cubic Casimir, which allows the $\mathbf{27}^3$ Yukawa coupling
and contains $\mathbf{16}\,\mathbf{16}\,\mathbf{10}$ and $\mathbf{10}\,\mathbf{10}\,\mathbf 1$. It gives
relations among three weights, such as $(2\hat e_1,-2)+(-2\hat e_1,-2)+(0,4)=0$, and
$(q,1)+(q',1)+(-q-q',-2)=0$ when $q$ and $q'$ agree in exactly one coordinate. Every weight lies in
one, so the removed weight is a combination of two others, and $d=2$ for every odd $p$. In the
$\mathbf{16}$ alone the $\mathbf{10}$ is a scalar and two couplings are needed, which is why the
logical errors there involve four fermions. 

The larger spinors give $d=3$, as we now show.
\begin{proposition}\label{prop:spinor}
Let $m=2k+1\ge5$, let $\Sigma_m\subset\{\pm1\}^m$ be the weights of a spinor of
$\mathfrak{so}(2m)$, the sign vectors with an even number of minus signs, and let
$\one=(1,\dots,1)\in\Sigma_m$. For every odd prime $p$, removing $\one$ gives a triorthogonal
$[[2^{m-1}-1,1,3]]_p$ code.
\end{proposition}

\begin{proof}
For the upper bound we exhibit three weights whose combination is $\one$. Pick
$u\in\{\pm1\}^{m-1}$ with an even number of minus signs and at least one sign of each kind. Then
$a=(1,u)$ and $b=(1,-u)$ lie in $\Sigma_m$, since $m-1$ is even, and so does $z=(1,-1,\dots,-1)$. None
of them is $\one$, and $a+b-z=\one$, so $d\le3$.

For the lower bound, every weight other than $\one$ has entries of both signs, since $-\one$ has the
wrong parity for odd $m$. So no single weight is proportional to $\one$, even modulo an odd prime.
Suppose that $\one=\alpha a+\beta b$ over $\Q$, with $a$ and $b$ distinct weights other than $\one$.
On the coordinates where $a$ and $b$ agree, $a$ is constant; on those where they differ, $a$ is
constant and $b=-a$. Both sets are nonempty, since $b\neq a$ and $b\neq-a$, the latter again by
parity. Whichever of $a$ and $b$ takes the same value on both sets is then constant, and so equals
$\pm\one$, a contradiction. Finally, the rank argument in the proof of Proposition~\ref{prop:reduction} carries
this to every odd prime $p$, since the minors it uses are nonzero $2\times2$ and $3\times3$ minors of $\pm1$
matrices, which are $\pm2$ and $\pm4$, so $p$ never divides them.
\end{proof}

For $m=5$ this is the distance in Theorem~\ref{thm:sm}, and for $m=7$ it gives a $[[63,1,3]]_p$ code.

\section*{Reproducibility}

The accompanying repository~\cite{LetterRepo} contains scripts that reproduce every computational
statement in the Letter and in this Supplement, and its documentation lists which script checks which
statement. They cover the charge sums and triorthogonality over $\Z$; the parameters of the code for
odd primes and its qubit form; exactness of the qutrit gate; the doubling identities; the minimal
logical errors and their classification; the codes obtained by removing several fermions; the spinor
codes; the exhaustive search and uniqueness; the automorphism groups; and the distillation figures.
The minimality, uniqueness and automorphism results were also checked independently in
Magma~\cite{Magma}.

\end{document}